# Symmetry Breaking and Transition to Robust Excitonic Topological Order in InAs/GaSb Bilayers

Xinghao Wang[1†], Wenfeng Zhang[1†], Yujiang Dong[1], Weiliang Qiao[1], Peizhe Jia[1], and Rui-Rui Du[1,2*]

[1]International Center for Quantum Materials, School of Physics, Peking University; Beijing 100871, China

[2]Hefei National Laboratory; Hefei 230088, China

[†]Both authors contribute equally to this work.

[*]Corresponding author. Email: rrd@pku.edu.cn

**ABSTRACT**

**Symmetry and topology are fundamental concepts deeply intertwined in various fields of physics, especially in the studies of quantum phases of matter. The critical role that Coulomb interactions play in symmetry breaking during topological transitions is a fundamental problem that has not been fully understood. Utilizing gated indium arsenide-gallium antimonide bilayers, we demonstrate that Coulomb interactions play a critical role in symmetry breaking and topological transitions. Whereas the quantum spin Hall insulator (QSHI) dominates the high-density regime, gating the system into the dilute regime enhances interlayer Coulomb interactions and leads to an emergent excitonic topological order (ETO) with spontaneous time-reversal-symmetry breaking. Moreover, applying a magnetic field drives a transition from the QSHI to the ETO accompanied by Coulomb-induced spin-rotation-symmetry breaking, which selects triplet electron-hole pairing in the lowest Landau levels. These results underscore an intricate interplay between symmetry and topology under Coulomb interactions in electron-hole bilayers.**

***Introduction*** - Many-body physics involving topology and correlations in quantum materials is at the forefront of condensed matter research. Recent years have seen the rise of topological insulators and topological superconductors [1-5]. Among semiconductor materials, inverted InAs/GaSb bilayers [6] which can simultaneously host electrons and holes in heterostructures, have been a fruitful playground for studying topology and interactions. For example, quantum spin Hall insulator (QSHI) with large bulk gap and quantized helical edge transport has been reported in InAs/GaSb and strained-layer InAs/InGaSb devices [7-17]. The QSHI here as described by the Bernevig-Hughes-Zhang model [4], is a single-particle $Z_2$ topological phase with time-reversal-symmetry (TRS)-protected helical edge states.

Both theoretical and experimental investigations on high-quality InAs/GaSb materials (as well as graphene and TMD materials), however, reveal that when the Coulomb interactions are taken into consideration, the system can also host novel many-body phases, such as excitonic insulator [18-25] and broken-symmetry states [25]. Excitonic condensation has been well-established in optically-pumped semiconductors [26,27]. While quantum Hall exciton superfluidity has been observed in GaAs/AlGaAs or graphene electronic bilayers [28-30], exciton superfluidity in electron-hole bilayers under zero magnetic field has only begun to be explored [31-35]. There exist critical questions concerning the many-body topological phase diagram in InAs/GaSb bilayers, such as how the two phases connect to each other, and under what conditions the topological phase transition occurs, and what the pairing symmetry of the excitonic ground state is. Here, we address these questions by systematic experimental investigations.

In the dilute limit, zero-momentum Bose-Einstein condensation (BEC) is the ground state in electron-hole bilayers [21]. The phase diagram of bilayer is greatly enriched when the carrier density is tuned into an imbalanced regime, resulting in finite-momentum excitonic states, such as density waves or Fulde-Ferrell-Larkin-Ovchinnikov (FFLO) states according to mean-field calculations [36, 37]. Surprisingly, in imbalanced InAs/GaSb bilayers we have recently observed a TRS breaking, gapped excitonic ground state [37] that is accompanied by a quantized, helical-like edge state with short coherence length [9,31]. Under a perpendicular magnetic field $B_\perp$, the edge transport shows a helical-like to chiral-like transition. Theoretical study beyond the mean-field level indicates that this TRS breaking state is associated with an intrinsic excitonic topological order (ETO) [37]. The density imbalance generates strong frustration, leading to a long-range entangled excitonic state in close analogy with the $\nu = 1/2$ bosonic fractional quantum Hall (FQH) state first proposed by Kalmeyer and Laughlin [38, 39]. Consistent with the fermionic FQH, the ETO theory beyond mean field can be mapped to composite fermion model [40], in which strong correlation generates emergent flux attachment. It is worth emphasizing that although there exists tangible evidence for bosonic FQH state observed in cold atoms [41], our ETO experiments have demonstrated a conceptually new bosonic platform in condensed matter systems. Whereas the QSHI is a TRS-protecting topological phase, the ETO is an intrinsic topological order [37]. Notably, we identify a new class of topological phase transition - between the QSHI and the ETO phase - that is continuously tunable through experimental parameters, *i.e.*, gate bias or perpendicular magnetic field $B_\perp$. The transition is deeply rooted in the interplay between topology and interlayer correlation in InAs/GaSb bilayers. Remarkably, our results indicate a triplet pairing symmetry for excitonic binding in the lowest Landau levels (LLL), therefore a spin current may be generated in the ETO state in InAs/GaSb bilayers.

We now turn to a brief introduction to the materials system. The degree of inversion, $E_g = E_1 - H_1$, where $E_1$ ($H_1$) is the ground state energy of the conduction (valence) band, is a crucial parameter in the phase diagram of InAs/GaSb bilayers [6]. The $E_g > 0$ corresponds to the normal insulator (NI) band; here we will focus on the inverted band, $E_g < 0$. In the QSHI phase, the hybridization of wavefunctions between the electron (InAs) and hole (GaSb) layers opens a minigap ($\Delta_{mini}$, see Fig. 1(b)) at specific wave vectors ($k_{cross}$), giving rise to a Kramers pair of spin-momentum-locked helical edge states [7]. $|E_g|$ can be quantified by the carrier density at the charge-neutral point (CNP), $n_{CNP} = k_{cross}^2/2\pi$. Experimentally, the phase diagram is divided by a "transition" density $n_{CNP,t} \sim 8 \times 10^{10} cm^{-2}$, above which the bilayers reach a deeply-inverted regime where the QSHI phase dominates [7-10,17].

On the opposite side is the shallowly-inverted regime, where the carrier density is relatively dilute and the plasmonic screening between electrons and holes is reduced, leading to the pairing instability of excitons [20-22]. Based on full Hartree-Fock calculations, Pikulin and Hyart first proposed [23,24] that in InAs/GaSb bilayers Coulomb interactions lead to a TRS-breaking phase when the CNPs are tuned from the NI to the QSHI regime. This excitonic state generally arises from the interplay between interlayer Coulomb interaction and interlayer tunneling. Under electron-hole density imbalance, such an excitonic state would be further driven towards the ETO phase proposed in Ref. [37]. The TRS-breaking effect results in a chiral edge state that consists of a pair of electron-like and hole-like edge channels, which are weakly bound by Coulomb attraction. Although at $B_\perp = 0$ the chiral-like edge state mimics the helical edge states of the QSHI, it does not require protection from TRS, thus it could display different transport behaviors from the QSHI under magnetic fields. The helical-like to chiral-like edge state transport from $B_\perp = 0$ to high $B_\perp$ is depicted in Fig. 1(c).

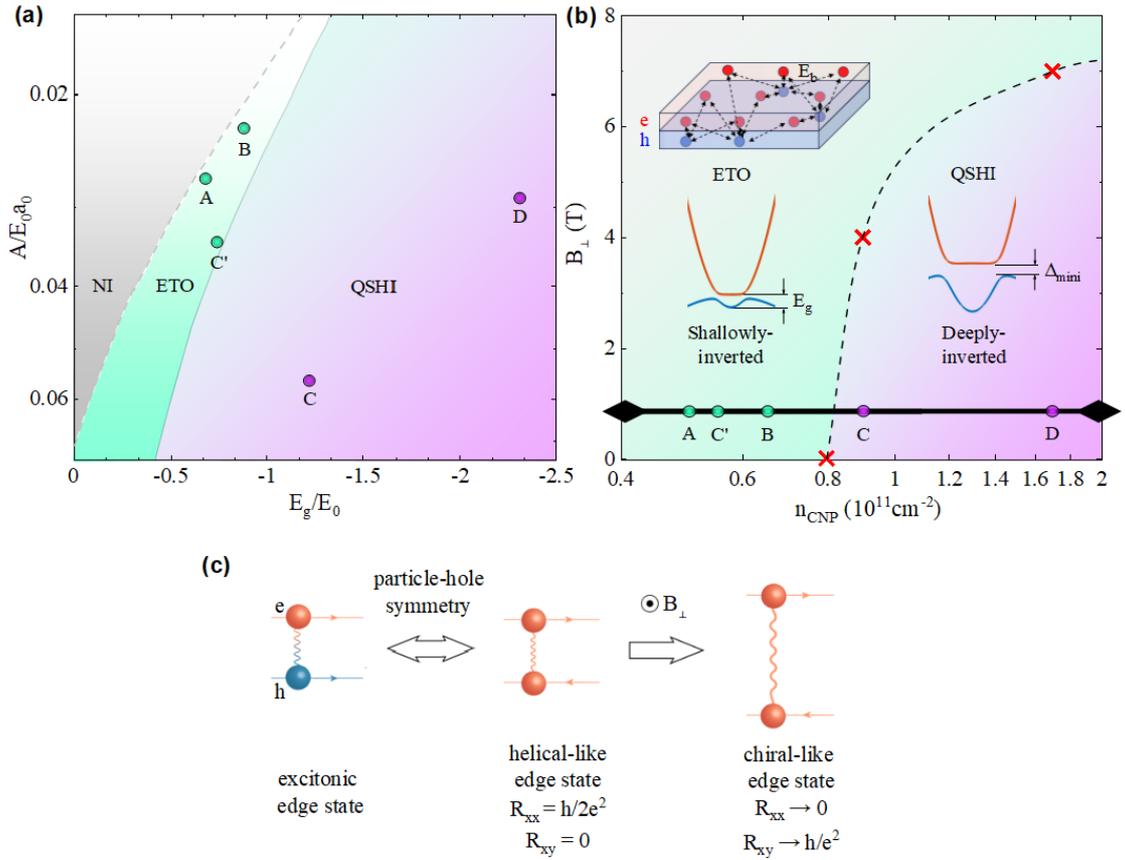

FIG. 1. (a) Empirical phase diagram at zero-magnetic field of inverted InAs/GaSb bilayers based on Ref. [23]. The four wafers (A-D) with zero back gate voltage are marked; C' marks wafer C

tuned by a back gate ($V_{bg} = 5V$) to a lower $n_{CNP}$. Parameters for each wafer were determined from measurements of $n_{CNP}$ and $\Delta_{mini}$ to derive $E_g$ and $A$ (see Supplemental Material (SM) Section III, IV). The phase boundaries are chosen to be the same as those in Fig.1(b) of Ref. [23]. (b) Experimental phase diagram of the ETO and QSHI with external magnetic field $B_\perp$. The dashed line approximately marks the phase boundary between ETO and QSHI; the red crosses represent the experimental data points presented in Fig. 2(d)-(f) and that in Fig. 4(a). (c) The ETO edge state transport evolves from helical-like at $B_\perp = 0$ to chiral-like with increasing $B_\perp$ (adapted from Ref. [37]).

Here we investigate three wafers grown by molecular beam epitaxy (MBE) (labeled B-D) with varying widths of InAs and GaSb layers. The wafers consist of InAs/GaSb composite quantum wells (QWs) capped with a 35 nm thick Al$_{0.7}$Ga$_{0.3}$Sb top barrier. The QW widths are 11/6 nm (wafer B), 12.5/5 nm (C), and 12.5/6 nm (D), respectively. Hall bars and Corbino devices were fabricated using lithography and wet etching. The longitudinal ($R_{xx}$) and Hall ($R_{xy}$) resistances at the respective CNPs, as well as the bulk energy gaps, were measured following standard low-temperature quantum transport procedures (for technical details, see Appendix A).

*Schematic phase diagram of topological phases -* According to the theoretical work [7, 23-25], InAs/GaSb bilayers exhibit topologically distinct phases determined by the bare band gap ($E_g = -\hbar^2 k_{cross}^2/2m^*$ when the energy band is inverted), the effective Rydberg constant $E_0$, and the tunneling parameter, $A \sim \Delta_{mini}/2k_{cross}$ (definition of $A$ can be referred to SM Section IV). Here, $m^* = m_e^* m_h^*/(m_e^* + m_h^*)$, where $m_e^* = 0.03 m_e$ and $m_h^* = 0.37 m_e$ represent the effective masses of electrons and holes, respectively ($m_e$ is the mass of free electrons). Both $E_g$ and $A$ are scaled in units of $E_0$ and $E_0 a_0$, where $E_0 = e^2/2\pi\epsilon_0\epsilon_r a_0 = 2\hbar^2/m^* a_0^2 = 6.36 meV$ and effective Bohr radius $a_0 = 29\ nm$. The dielectric constant $\epsilon_r = 15.4$ is taken from the average value for InAs and GaSb.

In the InAs/GaSb bilayers the ETO phase (green) emerges between the NI phase

(grey) and the QSHI phase (purple) (see Fig.1(a)). Among the four wafers (A-D), ETO signatures are observed in wafers A and B, in contrast to those of the QSHI in wafers C and D. In wafer C, a phase transition between the QSHI (C) and the ETO (C') is driven by reducing the CNP density via gate bias. For completeness, we also mark wafer A in Fig. 1(a, b), which shows archetypical ETO features as reported earlier in Ref. [37].

Fig. 1(b) presents a schematic experimental phase diagram, plotting $n_{CNP}$ vs. $B_\perp$. The application of $B_\perp$ generally increases the exciton binding energy; consequently, the phase boundary of the ETO phase shifts into the original QSHI regime. The insets illustrate the corresponding band structures of the phases, as well as the complex pairing mechanism proposed for the ETO.

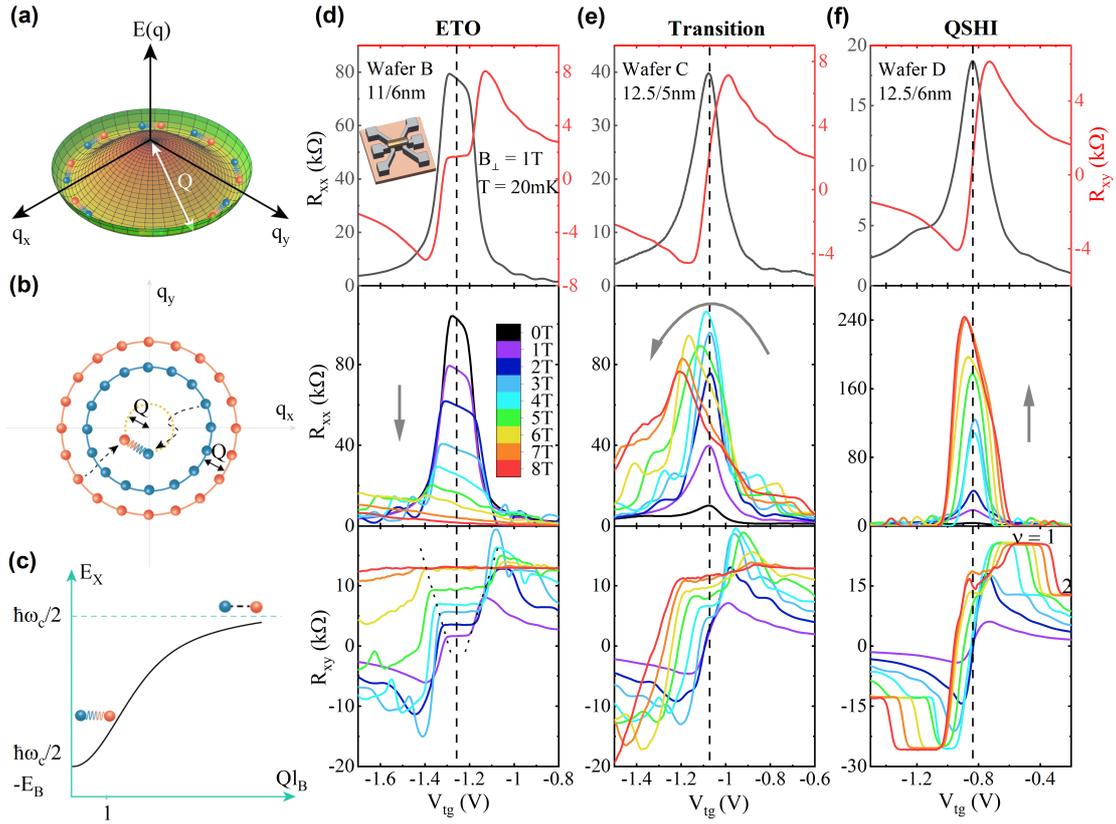

FIG. 2. (a, b) Schematic representation of a "moat band" [37]. (c) Theoretical dispersion for excitons under an out-of-plane magnetic field $B_\perp$ [42]. (d-f) Magneto-transport data highlighting the main differences between the ETO and QSHI states at $B_\perp = 1T$ for wafers B, C, and D; the most noticeable one is a Hall plateau for the ETO, signaling chiral edge transport. The evolutions

of $R_{xx}$ and $R_{xy}$, as $B_\perp$ varies from 0 T to 8 T, are shown in (d) for wafer B, (e) for C, and (f) for D. All data were measured at 20mK. The ETO state exhibits a decreasing $R_{xx}$ (marked by an arrow) and clear $R_{xy}$ plateaus as $B_\perp$ increases; the plateau width (marked by dotted lines) also increases. The QSHI exhibits opposite behaviors, *i.e.,* increasing $R_{xx}$ and no $R_{xy}$ plateaus. Wafer C exhibits a transition regime as characterized by $R_{xx}$ first increasing and then (around ~ 4 T) decreasing as a function of $B_\perp$; at the same time $R_{xy}$ plateaus appear at $B_\perp \sim 3\ T$.

***Topological phases and magnetic-field-induced phase transition -*** For clarity, we will first introduce the magneto-transport characteristics of the phases; the gate-tuned, zero-field transition from the QSHI to the ETO will be presented later in Fig. 4. Fig. 2 presents the data distinguishing the ETO state from the QSHI. As depicted in Fig. 2(a, b), the ETO occupies a "moat band" of radius $|Q| = \sqrt{2\pi}|\sqrt{n} - \sqrt{p}|$ which is highly degenerate for dilute excitons due to interaction and frustration [38]. The energy spectrum of magnetoexcitons is sketched in Fig. 2(c), which is appropriate for describing the electron-hole bilayer under $B_\perp$ [42]. In the high magnetic field limit, excitons are formed with a binding energy $E_B = \sqrt{\pi/2} e^2/\epsilon_0 \epsilon_r l_B \propto \sqrt{B_\perp}$, where $l_B = \sqrt{\hbar/eB_\perp}$ is the magnetic length. Under the density imbalance $n \neq p$, excitons in the moat band carry a momentum $P = \hbar Q$, with a dispersion described by modified Bessel functions of $Ql_B$. For the ETO phase, assuming a 2:1 imbalance, $n(p) = 1 \times 10^{11}(5 \times 10^{10}) cm^{-2}$, we obtain $Q = 5.6 \times 10^{-2} nm^{-1}$, giving $Ql_B = 1$ at $B_\perp \sim 2T$. With increasing $B_\perp$ the ETO phase can accommodate increasing $Q$, *i.e.,* an expanding moat band. This is consistent with the data showing increasing Hall plateau width in Fig. 2(d).

It should be noted that the ETO is driven by the correlation and frustration effects intrinsic to the moat band in the absence of an external magnetic field; the transport data measured here under $B_\perp$ are intended to reveal the internal structure of the edge state. On the other hand, the exact diagonalization results show that the moat band physics remains applicable at the LLL [37], where high degeneracy leads to divergence of the low-energy density of states, resulting in strong frustration in kinetic energy of bosons and preventing their condensation. In the moat band there remains residual electrons or holes after forming excitons, the interaction between the residual fermions and the bosons will not destroy the ETO. Instead, the fermion-boson coupling is in fact the primary mechanism that generates the ETO. The fermions are scattered by the excitons and form localized states at their Landau levels (LLs). Thus, the low-energy physics as well as transport properties are dominated by the edge states of the ETO state, which are measured in our experiments.

Fig. 2(d-f) respectively displays the $R_{xx}$ and $R_{xy}$ data for different wafers, which are qualitatively distinguishable. The ETO state (Fig. 2(d)) is characterized by a decreasing $R_{xx}$ (marked by an arrow) and clear $R_{xy}$ plateaus as $B_\perp$ increases; the plateau width (marked by dotted lines) also increases. Since the Hall plateau is a signature of chiral edge state transport, this observation is consistent with a magnetic-field-induced helical-like to chiral edge transport transition as depicted in Fig. 1(c). The plateau width increases with $B_\perp$, indicating a magnetic-field-induced widening of the density imbalance [37]. The QSHI state (Fig. 2(f)) exhibits opposite behaviors, *i.e.,* increasing $R_{xx}$ and the absence of $R_{xy}$ plateaus. The $R_{xx}$ signal contains two contributions: bulk resistance increases due to the magnetic-field-induced localization of residual bulk carriers, while the edge resistance increases because of gap-opening in the helical edge state [3]. Even at zero field, the shape of the $R_{xx}$ peak is distinct: for the ETO state the peak is beveled whereas for the QSHI it is sharply pointed.

Among all the wafers studied, wafer C stands out as unique due to its proximity to the phase transition boundary. With $n_{CNP} \sim 9 \times 10^{10} cm^{-2}$, it shows a typical QSHI behavior (Fig. 2(e)). Initially, $R_{xx}$ increases with $B_\perp$ in the range $B_\perp = 0 - 4\,T$ but decreases in the range of $B_\perp = 4 \sim 8\,T$. Simultaneously, the linear $R_{xy}$ at the CNP gradually develops into a plateau starting at $B_\perp = 3\,T$. These features suggest a transition from a QSHI state at zero field to an ETO state at high $B_\perp$. This transition can be partially understood according to the phase diagram in Fig. 1(b). A large $B_\perp$ enhances the excitonic correlation due to the increasing binding energy $E_B \propto \sqrt{B_\perp}$, competing with the interlayer tunneling.

***Large Gap energies of ETO and QSHI -*** The magnetic-field-induced topological phase transition from the QSHI to the NI insulator is expected in the absence of Coulomb interactions. On the other hand, anomalous conductance oscillations in Corbino devices have been reported in this regime but remain to be explained [43,44]. Recently, a topological transition from the QSHI to an excitonic state in the LLLs of InAs/GaSb bilayer is theoretically predicted [45]. Fig. 3(a) presents the LLL spectrum of inverted band structure, showing a level crossing between the *s*-electron state (spin up) and *p*-hole state (spin down) at a critical field $B_c$, while all other LLs disperse outwards due to band hybridization [45,46]. However, in the presence of interlayer Coulomb interactions [44,45], both space inversion symmetry and spin rotation symmetry are broken, and excitonic states can form near $B_c$. Our experiments here confirm a Coulomb-driven phase transition in InAs/GaSb bilayers. However, while the initial state is the QSHI, the final state is in fact the ETO - a long-range entangled topological order of excitons. Moreover, we have not found any signatures of gap closure or reopening during the QSHI-to-ETO transition (the gap values vs. $B_\perp$ for wafer C are plotted in Fig. 3(h)).

To systematically investigate the TRS-breaking ETO and the QSHI-to-ETO phase transition, we conducted gap energy measurements using Corbino devices. The bulk conductance of an intrinsic ETO in wafer B was measured at $B_\perp = 0 - 8\ T$ (Fig. 3(c)). Using Arrhenius plots, we extracted the energy gaps $\Delta$ at each $B_\perp$ in Fig. 3(d) and plotted them in Fig. 3(e). For wafer B, a large gap $\Delta(0T) \sim 48T$ is dominated by excitonic pairing and correlations, indicating that the ETO is formed even under a vanishingly small $B_\perp$ (see Appendix C). The ETO gap remains continuously open and is enhanced as $\Delta$ increases with $B_\perp$ from 48 K to 94 K. Note that a dip in $\Delta$ near 5 T is consistent with in-gap oscillations reported elsewhere [43].

For wafer C, the initial QSHI phase first exhibits a clear downturn in the energy gap around 1.5 T, followed by a sharp increase when $B_\perp > 4\ T$, reaching $\sim 44\ K$ at 8 T. By combining data from Hall bars and Corbino devices, we determined the edge resistance by subtracting the bulk contribution from the Hall bar data (Fig. 3(e, h)). Overall, we confirm that a topological transition from the QSHI to the ETO occurs near $B_\perp = 4\,T$.

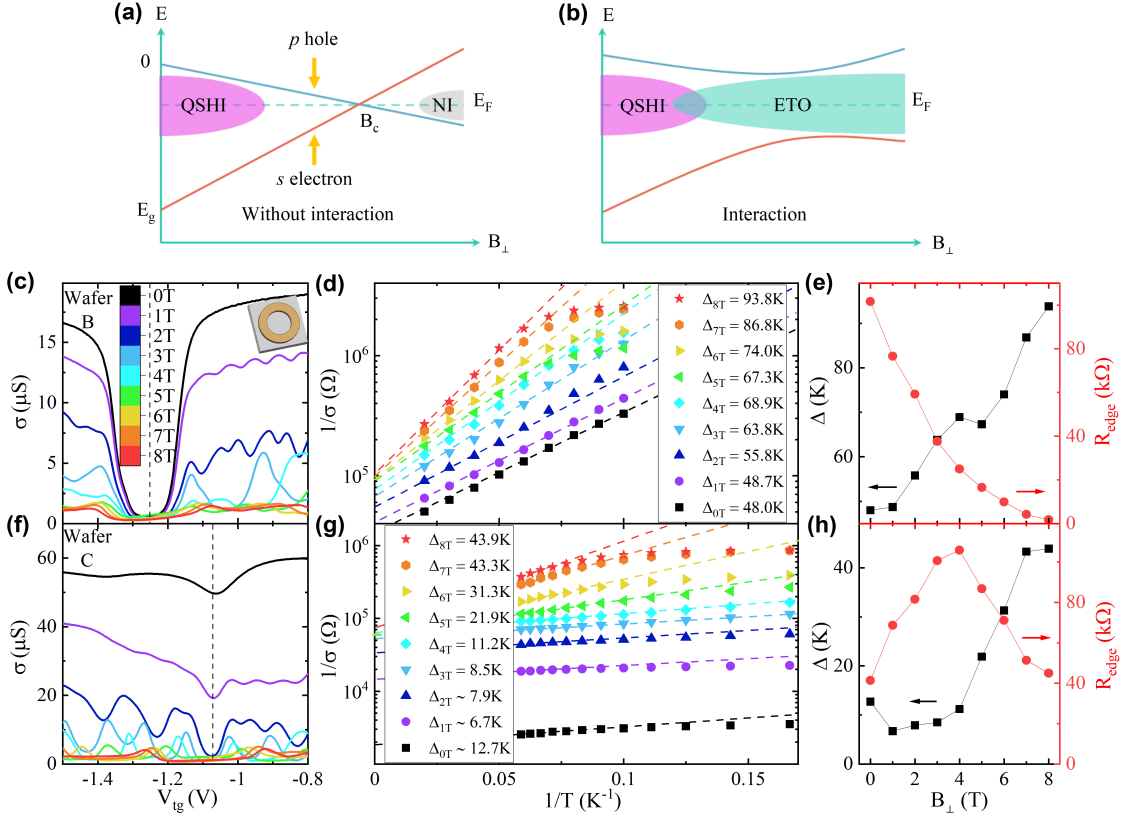

FIG. 3. (a) and (b) depict the LLL crossing (in the absence of interactions) and avoided LLL crossing (in the presence of interlayer Coulomb interaction). The bulk conductance of (c) wafer B and (f) wafer C is measured using Corbino devices at $300\,mK$ under a magnetic field $B_\perp$ ranging from 0 to 8 T; the data are presented in (d) for wafer B and (g) for wafer C, respectively. The measured energy gaps and the edge state resistance vs. $B_\perp$ are plotted in (e) for wafer B and (h) for wafer C. Notably, in panel (h) the inflection points in the energy gap and edge state resistance plots clearly mark the phase transition between the QSHI and the ETO phases.

***Tuning QSHI and ETO by gates*** - The topological phase diagram of wafer C, mapped by measuring the $R_{xx}$ as a function of top gate ($V_{tg}$) and back gate ($V_{bg}$) voltages at zero magnetic field, is shown in Fig. 4(a). Increasing the back gate voltage enhances the perpendicular electric field $E_z$, shifting the conduction band away from the valence band, reducing the degree of band inversion, and thereby transitioning the system from the QSHI (marked in Fig. 1(a, b) at point C, $n_{CNP} \sim 9 \times 10^{10} cm^{-2}$) to the ETO (at C', $n_{CNP} \sim 5.5 \times 10^{10} cm^{-2}$) phase. The two phases in Fig. 4(a) are marked by two high-resistance regions: the QSHI region typically exhibits an $R_{xx}$ peak of $9\ k\Omega$, while the ETO region shows a maximum $R_{xx}$ of approximately $50\ k\Omega$. The two regions meet at $V_{bg} = 1.5\ V$ and $V_{tg} = -1.0\ V$, where the $R_{xx}$ drops to about $7\ k\Omega$.

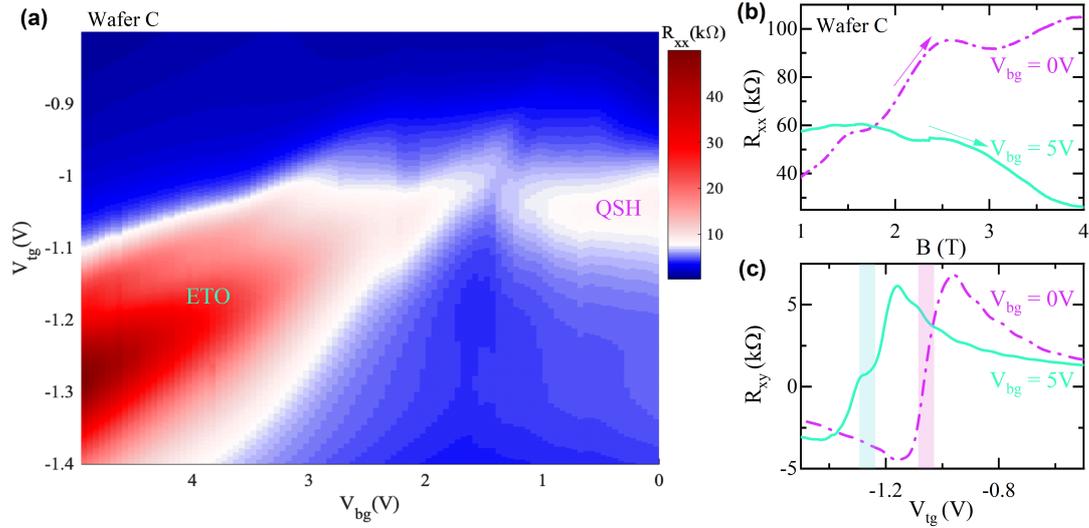

FIG. 4. (a) Phase diagram of wafer C, obtained by dual-gate tuning, revealing two topologically distinct phases: the ETO and the QSHI. Their characteristic transport features are illustrated in panels (b) and (c). In the ETO state ($V_{bg} = 5V$), $R_{xx}$ at the CNP decreases with increasing $B_\perp$, whereas in the QSHI state ($V_{bg} = 0V$), $R_{xx}$ increases. Additionally, the ETO state exhibits a Hall plateau in $R_{xy}$ at $B_\perp = 1\ T$, while the QSHI state does not. All data were measured at $20\ mK$.

As expected, they show opposite responses to $B_\perp$. Fig. 4(b) shows $R_{xx}$ versus $B_\perp$ at the CNPs, while Fig. 4(c) displays $R_{xy}$ traces at $B_\perp = 1T$ for $V_{bg} = 0V$ (QSHI) and $V_{bg} = 5V$ (ETO). In the QSHI phase (purple), $R_{xx}$ at the CNP increases with $B_\perp$, whereas in the ETO phase (green), it decreases. Moreover, $R_{xy}$ at the CNP in the QSHI phase varies linearly with $V_{tg}$, while in the ETO phase, it exhibits a developing Hall plateau.

***Discussions -*** The transition between the QSHI state and the ETO state represents a novel type of topological transition from a symmetry-protected state to a long-range entangled topological order, which is beyond the conventional Landau paradigm. The spontaneous TRS breaking in the ETO state is evidenced by the presence of a Hall plateau, interpreted as a signature of chiral edge state transport, even at near-zero magnetic fields. As $B_\perp$ increases and further breaks TRS, the ETO state becomes more robust. Our work thus demonstrates a remarkable experimental tunability of the topological transition across the boundary between short-range entangled and long-range entangled quantum phases in electron-hole bilayers. The present experimental results suggest that the ETO under the LLL condition favors triplet electron-hole pairing. It is conceivable to realize robust spin-triplet excitonic ground state (with a gap energy above 50 K at magnetic fields of order 1T), which would carry spin current. Spin-sensitive detection methods such as inverse spin Hall effect [47] should be performed on ETO devices for a systematic investigation. It would be natural to advance the investigation into even more diluted regime ($n_{CNP} < 5 \times 10^{10} cm^{-2}$), where Bose-Einstein condensation or FFLO state were theoretically predicted [21,36, 37]. With further refinement, the high-mobility InAs/GaSb electron-hole bilayers offer a clean material system for studying emergent phenomena such as Bose-Einstein condensation, bosonic FQH effect, excitonic crystals, or boson-fermion mixed states.

*Acknowledgements* - Helpful discussions with Rui Wang, Xincheng Xie, and Long Zhang are gratefully acknowledged. A portion of this work was carried out at the Synergetic Extreme Condition User Facility (SECUF). The work was funded by the Innovation Program for Quantum Science and Technology - National Science and Technology Major Project (Grant No. 2021ZD0302600), the National Key Research and Development Program of China (Grant No. 2024YFA1409002), and the Strategic Priority Research Program of Chinese Academy of Sciences (Grant No. XDB28000000).

# End Matter

*Appendix A: Experimental details* - The wafers are grown in a RIBER C21DZ MBE chamber equipped with valved crackers producing $Sb_2$ and $As_4$. Group III elements are evaporated by standard Knudsen cells. The substrates used are n-type GaSb (100) to match the lattice constants between the epitaxial layers and the substrate. The substrate is first degassed in the buffer chamber for 2 hours at 300°C and then transferred to the growth chamber. The substrate is then heated to 540°C under $Sb_2$ flux for 10 minutes to remove the native oxide. After oxide desorption, we first grow a 200 nm GaSb layer at 530°C to smooth the surface. Towards the end of this GaSb smoothing layer, an 800 nm $AlAs_ySb_{1-y}$ ($y = 0.08$) buffer layer is grown. Before the $Al_{0.7}Ga_{0.3}Sb$ bottom barrier, a short-period smoothing superlattice [10 × (2.5 nm GaSb + 2.5 nm AlSb)] is grown to improve the morphology and transport properties. Immediately before the growth of the $Al_{0.7}Ga_{0.3}Sb$ barrier, the substrate temperature is reduced from 530°C to 490°C, and maintained at this value for the remainder of the growth. The QW structure is flanked on both sides by outer $Al_{0.7}Ga_{0.3}Sb$ barrier layers and capped with a 3 nm-thick GaSb layer.

Hall bars (75 μm × 25 μm) and Corbino devices (with inner and outer radii of 550 μm and 600 μm, respectively) were fabricated using lithography and wet etching. Ohmic contacts were formed by soldering pure indium at 270°C for Hall bars, or by evaporating Ti/Au followed by annealing at 275°C for 20 min for Corbino devices. A conformal 100 nm-thick $HfO_2$ layer was grown *via* atomic layer deposition, serving as both the dielectric layer for the top gate and the passivation layer for the entire device. A 10 nm/40 nm Ti/Au gate was then deposited on top, while n-type GaSb substrate served as a back gate. The wafers B, C, D have identical layer structures but slightly varied thickness to accommodate required initial densities at zero back-gate voltage. Wafer C has a wider density turnability by a back gate, and can be turned to a lower $n_{CNP}$, as marked by C' ($V_{bg} = 5V$). The data point of wafer A is adapted from [37].

Measurements of $R_{xx}$ and $R_{xy}$ were conducted using an AC lock-in technique (17Hz) in a base temperature 300 mK ³He refrigerator with 9 T superconducting magnet, and in a base temperature 20 mK dilution refrigerator (18 T). Energy gaps of the Corbino samples were measured in the ³He refrigerator or a PPMS (9T). For Hall bar samples, to minimize the mixing of the $R_{xx}$ signal, $R_{xy}$ traces were anti-symmetrized with respect to $B_\perp$: $R_{xy} = (R_{xy,meas}(B_\perp) - R_{xy,meas}(-B_\perp))/2$, where $R_{xy,meas}$ is the measured Hall resistance.

To obtain the energy gap of each sample, we analyzed the Arrhenius plots shown in Fig. 3(b, e) and followed a standard procedure in quantum transport to deduce the energy gap: $\sigma \propto exp\,(-\Delta/2k_B T)$. The edge state resistance shown in Fig. 3(c, f) was calculated by subtracting the bulk resistance contribution measured in Corbino devices from the longitudinal resistance $R_{xx}$ measured in Hall bars. Since edge state resistance is parallelly connected to bulk resistance, we obtain $R_{edge} = 1/(1/R_{xx} - \sigma W/L)$, where $W$ and $L$ are the width and length of Hall bars, respectively.

*Appendix B: Recap moat band theory from Ref. [37]* - The moat band theory was first introduced in the context of cold atom systems, where the energy minima of the band form a degenerate loop in momentum space. Under such conditions, boson-boson interactions become dominant, giving rise to emergent gauge fluxes and favoring topological orders accompanied by anyonic excitations and spontaneous TRS breaking [48-51].

Excitons can spontaneously emerge in electron-hole bilayer systems, offering a promising platform for realizing moat-band physics. In systems with balanced electron-hole densities, condensation of zero-momentum excitons is favored, leading to topologically trivial exciton insulators. In contrast, under density imbalance, an electron may pair with different holes, resulting in numerous competing configurations with nearly degenerate energies, which is a manifestation of excitonic frustration [37]. For instance, in a minimal system containing one hole and three electrons, multiple pairing configurations with comparable energies emerge - analogous to frustrated quantum spins on a triangular lattice, which exhibit degenerate spin states.

In momentum space, density imbalance and frustration result in two concentric Fermi surfaces (Fig. 2(b)). Consequently, the formed excitons carry finite momenta lying on a loop satisfying $|q| = Q$, where $Q = \sqrt{2\pi}|\sqrt{n} - \sqrt{p}|$ is determined by carrier densities. These excitons exhibit equal propensity to condense along this loop, implying an infinite number of condensation channels. Interactions between excitons and excess particles substantially renormalize the excitonic dispersion, yielding a moat band structure, as depicted in Fig. 2(a). Furthermore, an effective exciton-exciton interaction of strength $U$ emerges, leading to the low-energy effective Hamiltonian:

$$H_b = \sum_r b_r^\dagger \left[ \frac{(|\hat{q}| - Q)^2}{2m_b} - \mu_b \right] b_r + U \sum_r n_{b,r} n_{b,r}, \tag{4}$$

where $\hat{q} = -i\nabla_r$, $m_b$ and $\mu_b$ denote the effective mass and chemical potential of excitons, respectively, and $n_{b,r} = b_r^\dagger b_r$ is the exciton particle number. It is important to note that electrons and holes form weakly paired states, and large binding energies are not a prerequisite. The emergent excitons are inherently strongly correlated owing to the flat dispersion along the moat $|q| = Q$.

The bosonic electron-hole pairs can be equivalently described as spinless fermions coupled to one quantum of flux [40]. The emergent flux acts as an effective magnetic field, giving rise to LLs. This provides a suitable wave function ansatz for the ground state of $H_b$: a fully filled LLL of composite fermions. This many-body state corresponds to a gapped, TRS-breaking topological order termed the ETO, with a chiral excitonic edge state. Using this ansatz, we find that the energy per particle is lower than that of the FFLO state [50] and the non-uniform condensation state [52] at relatively low exciton densities $n_b$. Compared with mean-field results, the ETO arises in the intermediate regime between the FFLO phase and the two-carrier region (QSHI), as a direct consequence of frustration effects.

*Appendix C: Signatures of spontaneous time-reversal-symmetry breaking* - Indirect transport evidence suggests spontaneous TRS breaking in the ETO state. A key characteristic of the ETO state is the decrease in $R_{xx}$ with increasing $B_\perp$, indicating that a chiral excitonic edge state gradually transforms into a chiral electronic edge state. The chirality of the excitonic edge state implies TRS breaking. As shown in Fig. 5(a), $R_{xx}$ for wafer B at the charge neutral point decreases with $B_\perp$ starting from a very low magnetic field around $B_\perp = 0.03$ T, suggesting that TRS breaking occurs near zero field. Furthermore, Fig. 5(b) shows an incipient Hall plateau for the ETO, indicating that chiral edge influences transport even at fields as small as 0.1 T.

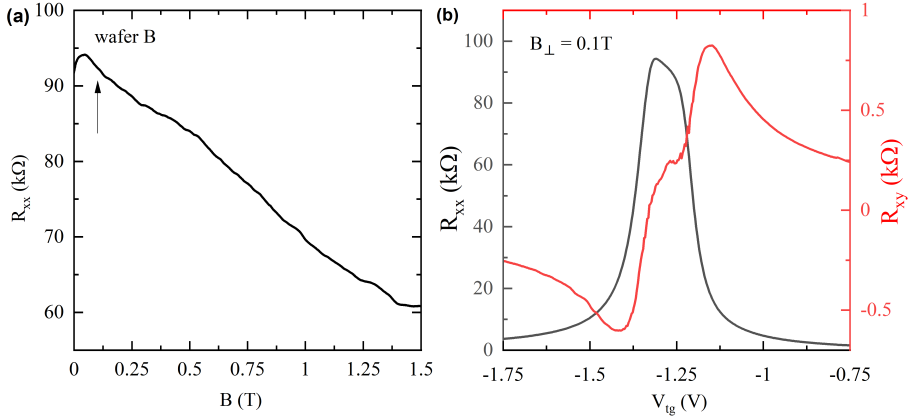

FIG. 5. (a) The longitudinal resistance of the ETO state decreases with magnetic field from $0.03\ T$, indicating the presence of a chiral excitonic edge state at sufficiently low magnetic field. (b) An incipient Hall plateau of the ETO state forms at $B_\perp = 0.1\ T$. The arrow in panel (a) marks the position of $B_\perp = 0.1\ T$.